\documentclass[journal]{IEEEtran}
\usepackage{graphicx,subfigure,amsmath}

\begin{document}

\title{Adaptive Synchronization of An Uncertain Complex Dynamical Network}

\author{Jin~Zhou,~
        Junan~Lu,~
        and~Jinhu~L\"u
\thanks{
        This work was supported by the National Natural Science Foundation
        of China under Grants 60304017, 20336040/B06, and 60574045, the
        National Key Basic Research and Development 973 Program of China
        under Grant 2003CB415200, and the Scientific Research Startup Special
        Foundation on Excellent PhD Thesis and Presidential Award of Chinese
        Academy of Sciences}
\thanks{J. Zhou and J. Lu are with the College of Mathematics and
Statistics, Wuhan University, Wuhan 430072, China.}
\thanks{J. L\"u is with the Key Laboratory of Systems and Control,
Institute of Systems Science, Academy of Mathematics and Systems
Science, Chinese Academy of Sciences, Beijing 100080, China
(Corresponding author e-mail: jhlu@iss.ac.cn).}}

\markboth{IEEE TRANSACTIONS ON AUTOMATIC CONTROL, 2006, 51(4):
652-656.}{Zhou\MakeLowercase{\textit{et al.}}: Adaptive
Synchronization of An Uncertain Complex Dynamical Network}

\maketitle

\begin{abstract}
This brief paper further investigates the locally and globally
adaptive synchronization of an uncertain complex dynamical network.
Several network synchronization criteria are deduced. Especially,
our hypotheses and designed adaptive controllers for network
synchronization are rather simple in form. It is very useful for
future practical engineering design. Moreover, numerical simulations
are also given to show the effectiveness of our synchronization
approaches.
\end{abstract}


\IEEEpeerreviewmaketitle

\section{Introduction}

\PARstart{O}{ver the past decade}, complex networks have been
intensively studied in various disciplines, such as social,
biological, mathematical, and engineering sciences [1-8]. A complex
network is a large set of interconnected nodes, where the nodes and
connections can be anything. Detailed examples are the World Wide
Web, Internet, communication networks, metabolic systems, food webs,
electrical power grids, and so on.

Recently, one of the interesting and significant phenomena in
complex dynamical networks is the synchronization of all dynamical
nodes in a network. In fact, synchronization is a kind of typical
collective behaviors and basic motions in nature. For example, the
synchronization of coupled oscillators can explain well many natural
phenomena. Furthermore, some synchronization phenomena are very
useful in our daily life, such as the synchronous transfer of
digital or analog signals in communication networks. Specifically,
synchronization in networks of coupled chaotic systems has received
a great deal of attention. Some synchronization criteria of two or
three Lorenz systems have been obtained in the literature. However,
it is often difficult to get the exact estimation of the coupling
coefficients since we do not know the exact boundary for most
chaotic systems. Up to now, we can only estimate the boundary of
very few chaotic systems [9-13], such as the Lorenz, Chen, L\"u
systems [14]. Moreover, we often know very little information on the
network structure, which makes network design very difficult. To
overcome these difficulties, an effectively adaptive synchronization
approach is proposed based on an uncertain complex dynamical network
model in this paper.

Slotine, Wang, and Rifai [16,17] further discussed the
synchronization of nonlinearly coupled continuous and hybrid
oscillators networks by using the contraction analysis approach
[18]. Bohacek and Jonckheere [19-20] proposed the so-called linear
dynamically varying method based on discrete time dynamical systems.
In the following, by using Lyapunov stability theory, several novel
locally and globally asymptotically stable network synchronization
criteria are deduced for an uncertain complex dynamical network.
Compared with some similar results [3,5,15], our sufficient
conditions for network synchronization are rather broad and the
controllers are very simple. It is very useful for future practical
engineering design. Moreover, our analysis method and network model
are very different from those of the above referenced literature
[16-20]. However, for some complex systems (e.g., biological
systems) with unknown couplings, our conditions are hard to be
verified. In fact, it is impossible to propose a universal
synchronization criterion for various complex networks since there
are many uncertain factors, such as network structures and coupling
mechanisms.

This paper is organized as follows. An uncertain complex dynamical
network model and several necessary hypotheses are given in Section
II. In Section III, locally and globally adaptive synchronization
criteria for uncertain complex dynamical networks are proposed. In
Section IV, a simple example is provided to verify the effectiveness
of the proposed method. Finally, conclusions are given in Section V.

\section{Preliminaries}

This section introduces an uncertain complex dynamical network
model and gives some preliminary definitions and hypotheses.

\subsection{An uncertain complex dynamical network model}

Consider an uncertain complex dynamical network consisting of $N$
identical nonlinear oscillators with uncertain nonlinear diffusive
couplings, which is described by
\begin{equation}\label{1}
    \dot{\textbf{x}}_i\,=\,\textbf{f}(\textbf{x}_i,\,t)\,+\,\textbf{h}_i
    (\textbf{x}_1,\,\textbf{x}_2,\,\cdots,\,\textbf{x}_N)\,+\,\textbf{u}_i,
\end{equation}
where $1\,\leq\,i\,\leq\,N$,
$\textbf{x}_i\,=\,(x_{i1},\,x_{i2},\,\cdots,\,x_{in})^T\in\,\textbf{R}^n$
is the state vector of the $i$th node,
$\textbf{f}\,:\,\Omega\,\times\,\textbf{R}^+\,\rightarrow\,\textbf{R}^n$
is a smooth nonlinear vector field, node dynamics is
$\dot{\textbf{x}}\,=\,\textbf{f}(\textbf{x},\,t)$,
$\textbf{h}_i\,:\,\Omega\,\times\,\cdots\,\times\,
\Omega\,\rightarrow\,\textbf{R}^n$ are unknown nonlinear smooth
diffusive coupling functions, $\textbf{u}_i\,\in\,\textbf{R}^n$ are
the control inputs, and the coupling-control terms satisfy
$\textbf{h}_i(\textbf{s},\,\textbf{s},\,\cdots,\,\textbf{s})
\,+\,\textbf{u}_i\,=\,\textbf{0}$, where $\textbf{s}$ is a
synchronous solution of the node system
$\dot{\textbf{x}}\,=\,\textbf{f}(\textbf{x},\,t)$.

\subsection{Preliminaries}

Network synchronization is a typical collective behavior. In the
following, a rigorous mathematical definition is introduced for
the concept of network synchronization.

\smallskip %
{\it Definition 1:} Let
$\textbf{x}_i(t\,;\,t_0,\,\textbf{X}_0)\;(1\,\leq\,i\,\leq\,N)$ be a
solution of the dynamical network (\ref{1}), where
$\textbf{X}_0\,=\,(\textbf{x}_1^0,\,\textbf{x}_2^0,\,\cdots,\,\textbf{x}_N^0)$,
$\textbf{f}\,:\,\Omega\,\times\,\textbf{R}^+\,\rightarrow\,\textbf{R}^n$,
and $\textbf{h}_i\,:\,\Omega\,\times\,\cdots\,\times\,
\Omega\,\rightarrow\,\textbf{R}^n\,(1\,\leq\,i\,\leq\,N)$ are
continuously differentiable, $\Omega\,\subseteq\,\textbf{R}^n$. If
there is a nonempty subset $\Lambda\,\subseteq\,\Omega$, with
$\textbf {x}_i^0\in\,\Lambda\,(1\,\leq\,i\,\leq\,N)$, such that
$\textbf {x}_i(t\,;\,t_0,\,\textbf{X}_0)\,\in\,\Omega$ for all
$t\,\geq\,t_0$, $1\,\leq\,i\,\leq\,N$, and
\begin{equation}\label{2}
\lim\limits_{t\rightarrow\infty}\left\|\textbf
{x}_i(t\,;\,t_0,\,\textbf{X}_0)-\textbf{s}(t\,;\,t_0,\,\textbf
{x}_0)\right\|_2\,=\,\textbf{0}\quad1\leq\,i\leq\,N
\end{equation}
where $\textbf{s}(t\,;\,t_0,\,\textbf{x}_0)$ is a solution of the
system $\dot{\textbf{x}}\,=\,\textbf{f}(\textbf{x},\,t)$ with
$\textbf{x}_0\,\in\,\Omega$, then the dynamical network (\ref{1}) is
said to realize {\it synchronization} and
$\Lambda\,\times\,\cdots\,\times\,\Lambda$ is called the {\it region
of synchrony} for the dynamical network (\ref{1}).%
\smallskip

Hereafter, denote $\textbf{s}(t\,;\,t_0,\,\textbf{x}_0)$ as
$\textbf{s}(t)$. Then
$\textbf{S}(t)\,=\,(\textbf{s}^T(t),\,\textbf{s}^T(t),\,\cdots,\,\textbf{s}^T(t))^T$
is a synchronous solution of uncertain dynamical network (\ref{1})
since it is a diffusive coupling network. Here, $\textbf{s}(t)$ can
be an equilibrium point, a periodic orbit, an aperiodic orbit, or a
chaotic orbit in the phase space.

Define error vector
\begin{equation}\label{3}
    \textbf{e}_i(t)\,=\,\textbf{x}_i(t)\,-\,\textbf{s}(t),\quad
                      1\,\leq\,i\,\leq\,N\,.
\end{equation}
Then the objective of controller $\textbf{u}_i$ is to guide the
dynamical network (\ref{1}) to synchronize. That is,
\begin{equation}\label{4}
    \lim\limits_{t\rightarrow+\infty}\|\textbf{e}_i(t)\|_2\,=\,0\,,
    \quad 1\,\leq\,i\,\leq\,N\,.
\end{equation}

Since $\dot{\textbf{s}}\,=\,\textbf{f}(\textbf{s},\,t)$, from
network (\ref{1}), we have
\begin{equation}\label{5}
    \dot{\textbf{e}}_i\,=\,\bar{\textbf{f}}(\textbf{x}_i,\,
    \textbf{s},\,t)\,+\,\bar{\textbf{h}}_i(\textbf{x}_1,\,\textbf{x}_2,
    \,\cdots,\,\textbf{x}_N,\,\textbf{s})\,+\,\textbf{u}_i\,,
\end{equation}
where $1\,\leq\,i\,\leq\,N$, $\bar{\textbf{f}}(\textbf{x}_i,\,
\textbf{s},\,t)\,=\,\textbf{f}(\textbf{x}_i,\,t)\,-\,\textbf{f}
(\textbf{s},\,t)$, $\bar{\textbf{h}}_i(\textbf{x}_1,\,
\textbf{x}_2,\,\cdots,\,\textbf{x}_N,\,\textbf{s})\,=\,\textbf{h}_i
(\textbf{x}_1,\,\textbf{x}_2,\,\cdots,\,\textbf{x}_N)\,-\,\textbf{h}_i
(\textbf{s},\,\textbf{s},\,\cdots,\,\textbf{s})$.

In the following, we give several useful hypotheses.

\smallskip %
{\it Hypothesis 1:} (H1) Assume that there exists a nonnegative
constant $\alpha$ satisfying
$\|\mbox{D}\textbf{f}(\textbf{s},\,t)\|_2\,=\,\|\textbf{A}(t)\|_2\,\leq\,\alpha$,
where $\textbf{A}(t)$ is the Jacobian of
$\textbf{f}(\textbf{s},\,t)$.

\smallskip %
{\it Hypothesis 2:} (H2) Suppose that there exist nonnegative
constants $\gamma_{ij}\,(1\,\leq\,i,\,j\,\leq\,N)$ satisfying
$\|\bar{\textbf{h}}_i(\textbf{x}_1,\,\textbf{x}_2,\,
\cdots,\,\textbf{x}_N,\,\textbf{s})\|_2\,\leq\,\sum\limits_{j=1}^{N}\gamma_{ij}
\|\textbf{e}_j\|_2$ for $1\,\leq\,i\,\leq\,N$.

\smallskip %
{\it Remark 1:} If H1 holds, then we get
$\left\|\frac{\textbf{A}(t)\,+\,\textbf{A}^T(t)}{2}\right\|_2\,\leq\,\alpha$.

\section{Adaptive synchronization of an uncertain complex dynamical network}

This section discusses the local synchronization and global
synchronization of the uncertain complex dynamical network
(\ref{1}). Several network synchronization criteria are given.

\subsection{Local Synchronization}

Linearizing error system (\ref{5}) around zero gives
\begin{equation}\label{6}
    \dot{\textbf{e}}_i\,=\,\textbf{A}(t)\textbf{e}_i(t)\,+\,\bar{\textbf{h}}_i
    (\textbf{x}_1,\,\textbf{x}_2,\,\cdots,\,\textbf{x}_N,\,\textbf{s})\,+\,\textbf{u}_i\,,
\end{equation}
where $1\,\leq\,i\,\leq\,N$ and recall that
$\textbf{A}(t)\,=\,\mbox{D}\textbf{f} (\textbf{s},\,t)$ is the
Jacobian of $\textbf{f}$ evaluated at
$\textbf{x}\,=\,\textbf{s}(t)$.

Based on H1 and H2, a network synchronization criterion is deduced
as follows.

\smallskip %
{\it Theorem 1:} Suppose that H1 and H2 hold. Then the synchronous
solution $\textbf{S}(t)$ of uncertain dynamical network (\ref{1}) is
locally asymptotically stable under the adaptive controllers
\begin{equation}\label{7}
    \textbf{u}_i\,=\,-d_i\textbf{e}_i,\quad 1\,\leq\,i\,\leq\,N
\end{equation}
and updating laws
\begin{equation}\label{8}
    \dot{d}_i\,=\,k_i\,\textbf{e}_i^T\textbf{e}_i\,=\,k_i\|\textbf{e}_i\|_2^2
    \,,\quad 1\,\leq\,i\,\leq\,N\,,
\end{equation}
where $k_i\,(1\,\leq\,i\,\leq\,N)$ are positive constants.
\smallskip %

{\it Proof:} Define a Lyapunov candidate as follows:
\begin{equation}\label{9}
    V\,=\,\frac{1}{2}\sum\limits_{i\,=\,1}^N\textbf{e}^T_i\textbf{e}_i
    \,+\,\frac{1}{2}\sum\limits_{i\,=\,1}^N\frac{(d_i\,-\,\hat{d}_i)^2}{k_i}\,,
\end{equation}
where $\hat{d}_i\,(1\,\leq\,i\,\leq\,N)$ are positive constants to
be determined. Thus one gets
$$
\begin{array}{rcl}
  \dot{V} & = &
  \frac{1}{2}\sum\limits_{i\,=\,1}^N(\dot{\textbf{e}}^T_i\textbf{e}_i\,
  +\,\textbf{e}^T_i\dot{\textbf{e}}_i)\,+\,\sum\limits_{i\,=\,1}^N
  \frac{(d_i\,-\,\hat{d}_i)\dot{d}_i}{k_i} \\
          & = & \sum\limits_{i\,=\,1}^N\textbf{e}^T_i\left(\frac{\textbf{A}(t)
                \,+\,\textbf{A}^T(t)}{2}\,-\,d_i\textbf{I}_n\right)\textbf{e}_i \\
          &   & +\,\sum\limits_{i\,=\,1}^N\textbf{e}^T_i\bar{\textbf{h}}_i(\textbf{x}_1
                ,\,\textbf{x}_2,\,\cdots,\,\textbf{x}_N,\,\textbf{s})\,+\,
                \sum\limits_{i\,=\,1}^N(d_i\,-\,\hat{d}_i)\textbf{e}_i^T\textbf{e}_i \\
          & \leq & \sum\limits_{i\,=\,1}^N\textbf{e}^T_i\left(\frac{\textbf{A}(t)
                   \,+\,\textbf{A}^T(t)}{2}\,-\,\hat{d}_i\textbf{I}_n\right)\textbf{e}_i \\
          &      & +\,\sum\limits_{i\,=\,1}^N\sum\limits_{j\,=\,1}^N\,\gamma_{ij}
                   \|\textbf{e}_i\|_2\|\textbf{e}_j\|_2 \\
          & \leq & \sum\limits_{i\,=\,1}^N(\alpha\,-\,\hat{d}_i)\|\textbf{e}_i\|_2^2
                   \,+\,\sum\limits_{i\,=\,1}^N\sum\limits_{j\,=\,1}^N\gamma_{ij}
                   \|\textbf{e}_i\|_2\|\textbf{e}_j\|_2 \\
          & = & \textbf{e}^T(\bf{\Gamma}\,+\,\mbox{diag}\{\alpha\,-\,\hat{\mbox{d}}_1,\,
                \alpha\,-\,\hat{\mbox{d}}_2,\,\cdots,\,\alpha\,-\,\hat{\mbox{d}}_N\})\textbf{e}\,, \\
\end{array}
$$
where
$\textbf{e}\,=\,(\|\textbf{e}_1\|_2,\,\|\textbf{e}_2\|_2,\,\cdots,\,\|\textbf{e}_N\|_2)^T$
and $\bf{\Gamma}\,=\,(\gamma_{ij})_{N\,\times\,N}$.

Since $\alpha$ and $\gamma_{ij}\,(1\,\leq\,i,\,j\,\leq\,N)$ are
nonnegative constants, one can select suitable constants
$\hat{d}_i\,(1\,\leq\,i\,\leq\,N)$ to make
$\bf{\Gamma}\,+\,\mbox{diag}\{\alpha\,-\,\hat{\mbox{d}}_1,\,\alpha\,
-\,\hat{\mbox{d}}_2,\,\cdots,\,\alpha\,-\,\hat{\mbox{d}}_N\}$ a
negative definite matrix. Thus it follows that the error vector
$\bf{\eta}\,=\,(\textbf{e}_1^T,\,\textbf{e}_2^T,\,\cdots,\,
\textbf{e}_N^T)^T\,\rightarrow\,0$ as $t\,\rightarrow\,+\infty$.
That is, the synchronous solution $\textbf{S}(t)$ of uncertain
dynamical network (\ref{1}) is locally asymptotically stable under
the adaptive controllers (\ref{7}) and updating laws (\ref{8}).

The proof is thus completed.

Assume that the coupling of network (\ref{1}) is linear satisfying
$\textbf{h}_i(\textbf{x}_1,\,\textbf{x}_2,\,\cdots,\,\textbf{x}_N)\,=\,
\sum\limits_{j=1}^N\,b_{ij}\textbf{x}_j$ for $1\,\leq\,i\,\leq\,N$,
where $b_{ij}\,(1\,\leq\,i,\,j\,\leq\,N)$ are constants. Then the
uncertain network (\ref{1}) is recasted as follows:
\begin{equation}\label{10}
    \dot{\textbf{x}}_i\,=\,\textbf{f}(\textbf{x}_i,\,t)\,+\,
    \sum\limits_{j\,=\,1}^N\,b_{ij}\textbf{x}_j\,+\,\textbf{u}_i\,,
    \quad 1\,\leq\,i\,\leq\,N\,.
\end{equation}

For linear coupling, H2 is naturally satisfied. Thus one gets the
following corollaries.

\smallskip %
{\it Corollary 1:} Suppose that H1 holds. Then the synchronous
solution $\textbf{S}(t)$ of the uncertain dynamical network
(\ref{10}) is locally asymptotically stable under the adaptive
controllers (\ref{7}) and updating laws (\ref{8}).

Moreover, for the coupling scheme
$\textbf{h}_i(\textbf{x}_1,\,\textbf{x}_2,\,\cdots,\,\textbf{x}_N)\,=\,
\sum\limits_{j=1}^N\,b_{ij}\textbf{p}(\textbf{x}_j)$ with
$1\,\leq\,i\,\leq\,N$, where $b_{ij}\,(1\,\leq\,i,\,j\,\leq\,N)$ are
constants satisfying $\sum\limits_{j=1}^N\,b_{ij}\,=\,0$ for
$1\,\leq\,i\,\leq\,N$ and
$\|\mbox{D}\textbf{p}({\bf\xi})\|_2\,\leq\,\delta$ for
${\bf\xi}\,\in\,\Omega$, the network (\ref{1}) is rewritten as
follows:
\begin{equation}\label{11}
    \dot{\textbf{x}}_i\,=\,\textbf{f}(\textbf{x}_i,\,t)\,+\,
    \sum\limits_{j\,=\,1}^N\,b_{ij}\textbf{p}(\textbf{x}_j)\,+\,\textbf{u}_i\,,
    \quad 1\,\leq\,i\,\leq\,N\,.
\end{equation}

If H1 holds, then one has
$$
\begin{array}{rcl}
    \|\bar{\textbf{h}}_i(\textbf{x}_1,\,\textbf{x}_2,\,\cdots,\,
    \textbf{x}_N,\,\textbf{s})\|_2 & = &
    \sum\limits_{j\,=\,1}^N\,|b_{ij}|\left\|\textbf{p}(\textbf{x}_j)
    \,-\,\textbf{p}(\textbf{s})\right\|_2 \\
    & \leq & \sum\limits_{j\,=\,1}^N\,\delta\,|b_{ij}|\|\textbf{e}_j\|_2 \\
\end{array}
$$
for $1\,\leq\,i\,\leq\,N$. That is, H2 holds and one gets the
following corollary.

\smallskip %
{\it Corollary 2:} Assume that H1 holds. Then, under the adaptive
controllers (\ref{7}) and updating laws (\ref{8}), the synchronous
solution $\textbf{S}(t)$ of the uncertain dynamical network
(\ref{11}) is locally asymptotically stable.

In the following subsection, we discuss the global synchronization
case.

\subsection{Global synchronization}

This section presents two global network synchronization criteria.

Rewrite node dynamics
$\dot{\textbf{x}}_i\,=\,\textbf{f}(\textbf{x}_i,\,t)$ as
$\dot{\textbf{x}}_i\,=\,\textbf{B}\textbf{x}_i(t)\,+\,\textbf{g}(\textbf{x}_i,\,t)$,
where $\textbf{B}\,\in\,\textbf{R}^{n\,\times\,n}$ is a constant
matrix and
$\textbf{g}\,:\,\Omega\,\times\,\textbf{R}^+\,\rightarrow\,\textbf{R}^n$
is a smooth nonlinear function. Thus network (\ref{1}) is described
by
\begin{equation}\label{12}
    \dot{\textbf{x}}_i\,=\,\textbf{B}\textbf{x}_i(t)\,+\,\textbf{g}
    (\textbf{x}_i,\,t)\,+\,\textbf{h}_i(\textbf{x}_1,\,\textbf{x}_2,
    \,\cdots,\,\textbf{x}_N)\,+\,\textbf{u}_i,
\end{equation}
where $1\,\leq\,i\,\leq\,N$. Similarly, one can get the error
system
\begin{equation}\label{13}
    \dot{\textbf{e}}_i\,=\,\textbf{B}\textbf{e}_i(t)\,+\,\bar{\textbf{g}}
    (\textbf{x}_i,\,\textbf{s},\,t)\,+\,\bar{\textbf{h}}_i(\textbf{x}_1,
    \,\textbf{x}_2,\,\cdots,\,\textbf{x}_N,\,\textbf{s})\,+\,\textbf{u}_i,
\end{equation}
where $1\,\leq\,i\,\leq\,N$ and
$\bar{\textbf{g}}(\textbf{x}_i,\,\textbf{s},\,t)\,=\,\textbf{g}(
\textbf{x}_i,\,t)\,-\,\textbf{g}(\textbf{s},\,t)$.

\smallskip %
{\it Hypothesis 3:} (H3) Suppose that there exists a nonnegative
constant $\mu$ satisfying
$\|\bar{\textbf{g}}(\textbf{x}_i,\,\textbf{s},\,t)\|_2\,\leq\,\mu\,
\|\textbf{e}_i\|_2$.

Then one can get the following global network synchronization
criterion.

\smallskip %
{\it Theorem 2:} Suppose that H2 and H3 hold. Then the synchronous
solution $\textbf{S}(t)$ of uncertain dynamical network (\ref{1}) is
globally asymptotically stable under the adaptive controllers
\begin{equation}\label{14}
    \textbf{u}_i\,=\,-d_i\textbf{e}_i,\quad 1\,\leq\,i\,\leq\,N
\end{equation}
and updating laws
\begin{equation}\label{15}
    \dot{d}_i\,=\,k_i\,\textbf{e}_i^T\textbf{e}_i\,=\,k_i\|\textbf{e}_i\|_2^2
    \,,\quad 1\,\leq\,i\,\leq\,N\,,
\end{equation}
where $k_i\,(1\,\leq\,i\,\leq\,N)$ are positive constants.
\smallskip %

{\it Proof:} Since $\textbf{B}$ is a given constant matrix, there
exists a nonnegative constant $\beta$ such that
$\|\textbf{B}\|_2\,\leq\,\beta$. It follows that
$\left\|\frac{\textbf{B}\,+\,\textbf{B}^T}{2}\right\|_2\,\leq\,\beta$.

Similarly, construct Lyapunov function (\ref{9}), then one has
$$
\begin{array}{rcl}
  \dot{V} & = &
  \sum\limits_{i\,=\,1}^N\textbf{e}^T_i\left(\frac{\textbf{B}
                \,+\,\textbf{B}^T}{2}\,-\,\hat{d}_i\textbf{I}_n\right)\textbf{e}_i
                +\,\sum\limits_{i\,=\,1}^N\textbf{e}^T_i\bar{\textbf{g}}_i(\textbf{x}_i
                ,\,\textbf{s},\,t) \\
          &   & +\,\sum\limits_{i\,=\,1}^N\textbf{e}^T_i\bar{\textbf{h}}_i
                (\textbf{x}_1,\,\textbf{x}_2,\,\cdots,\,\textbf{x}_N,\,\textbf{s}) \\
          & \leq & \sum\limits_{i\,=\,1}^N(\beta\,+\,\mu\,-\,\hat{d}_i)\|\textbf{e}_i\|_2^2
                   \,+\,\sum\limits_{i\,=\,1}^N\sum\limits_{j\,=\,1}^N\gamma_{ij}
                   \|\textbf{e}_i\|_2\|\textbf{e}_j\|_2 \\
          & = & \textbf{e}^T(\bf{\Gamma}\,+\,\mbox{diag}\{\beta\,+\,\mu\,-\,\hat{\mbox{d}}_1,\,
                \cdots,\,\beta\,+\,\mu\,-\,\hat{\mbox{d}}_N\})\textbf{e}\,,\\
\end{array}
$$
where
$\textbf{e}\,=\,(\|\textbf{e}_1\|_2,\,\|\textbf{e}_2\|_2,\,\cdots,\,\|\textbf{e}_N\|_2)^T$
and $\bf{\Gamma}\,=\,(\gamma_{ij})_{N\,\times\,N}$.

Since $\beta,\,\mu$ and $\gamma_{ij}\,(1\,\leq\,i,\,j\,\leq\,N)$ are
nonnegative constants, one can select suitable constants
$\hat{d}_i\,(1\,\leq\,i\,\leq\,N)$ to make
$\bf{\Gamma}\,+\,\mbox{diag}\{\beta\,+\,\mu\,-\,\hat{\mbox{d}}_1,\,\beta\,
+\,\mu\,-\,\hat{\mbox{d}}_2,\,\cdots,\,\beta\,+\,\mu\,-\,\hat{\mbox{d}}_N\}$
a negative definite matrix. Then the error vector
$\bf{\eta}\,=\,(\textbf{e}_1^T,\,\textbf{e}_2^T,\,\cdots,\,
\textbf{e}_N^T)^T\,\rightarrow\,0$ as $t\,\rightarrow\,+\infty$.
That is, the synchronous solution $\textbf{S}(t)$ of uncertain
dynamical network (\ref{1}) is globally asymptotically stable under
the adaptive controllers (\ref{14}) and updating laws (\ref{15}).

This completes the proof.

Similarly, one gets the following two corollaries of global
network synchronization.

\smallskip %
{\it Corollary 3:} Suppose that H3 holds. Then the synchronous
solution $\textbf{S}(t)$ of uncertain linearly coupled dynamical
network (\ref{10}) is globally asymptotically stable under the
adaptive controllers (\ref{14}) and updating laws (\ref{15}).

\smallskip %
{\it Corollary 4:} Suppose that H3 holds. Then the synchronous
solution $\textbf{S}(t)$ of uncertain dynamical network (\ref{11})
is globally asymptotically stable under the adaptive controllers
(\ref{14}) and updating laws (\ref{15}).
\smallskip %

{\it Proof.} According to (\ref{11}), one has
$$
\begin{array}{rcl}
  \|\bar{\textbf{h}}_i(\textbf{x}_1,\textbf{x}_2,\cdots,
  \textbf{x}_N,\textbf{s})\|_2
  & = & \left\|\sum\limits_{j=1}^N\,b_{ij}(\textbf{Be}_j+
  \bar{\textbf{g}}_i(\textbf{x}_j,\textbf{s},t))\right\|_2 \\
  & \leq & \sum\limits_{j\,=\,1}^N\,b_{ij}(\|\textbf{B}\|\,
  +\,\mu)\|\textbf{e}_j\|\,; \\
\end{array}
$$
thus H2 holds. Therefore, from Theorem 2, the synchronous solution
$\textbf{S}(t)$ of network (\ref{11}) is globally asymptotically
stable under the adaptive controllers (\ref{14}) and updating laws
(\ref{15}).

The proof is thus completed.

\smallskip %
{\it Hypothesis 4:} (H4) Assume that $\textbf{g}(\textbf{x},\,t)$
satisfies the Lipschitz condition. That is, there exists a positive
constant $\kappa$ satisfying
$\|\textbf{g}(\textbf{x},\,t)\,-\,\textbf{g}(\textbf{y},\,t)\|\,
\leq\,\kappa\|x\,-\,y\|$, where $\kappa$ is the Lipschitz constant.

Obviously, H4 implies H3. Now one has the following
synchronization criterion.

\smallskip %
{\it Theorem 3:} Suppose that H2 and H4 hold. Then the synchronous
solution $\textbf{S}(t)$ of uncertain dynamical network (\ref{1}) is
globally asymptotically stable under the adaptive controllers
(\ref{14}) and updating laws (\ref{15}).

\section{An example}

This section presents an example to show the effectiveness of the
above synchronization criteria.

Consider a dynamical network consisting of $50$ identical Lorenz
systems. Here, node dynamics is described by

$$
\left(\begin{array}{c}
  \dot{x}_{i1} \\
  \dot{x}_{i2} \\
  \dot{x}_{i3} \\
\end{array}\right)\,=\,\textbf{A}\left(\begin{array}{c}
  x_{i1} \\
  x_{i2} \\
  x_{i3} \\
\end{array}\right)\,+\,\left(\begin{array}{c}
  0 \\
  -x_{i1}x_{i3} \\
  x_{i1}x_{i2} \\
\end{array}\right)\,,
$$
where
$$
\textbf{A}\,=\,\left(\begin{array}{ccc}
  -a & a & 0 \\
  c  & -1 & 0 \\
  0  & 0  & -b \\
\end{array}\right)\,,
$$
$a\,=\,10,\,b\,=\,\frac{8}{3},\,c\,=\,28$ and
$1\,\leq\,i\,\leq\,50$. And the networked system is defined as
follows:
\begin{equation}\label{16}
    \begin{array}{rcl}
    \left(\begin{array}{c}
  \dot{x}_{i1} \\
  \dot{x}_{i2} \\
  \dot{x}_{i3} \\
\end{array}\right) & = & \textbf{A}\left(\begin{array}{c}
  x_{i1} \\
  x_{i2} \\
  x_{i3} \\
\end{array}\right)\,+\,\left(\begin{array}{c}
  0 \\
  -x_{i1}x_{i3} \\
  x_{i1}x_{i2} \\
\end{array}\right) \\
    &  & +\left(\begin{array}{c}
  f_1(\textbf{x}_i)-2f_1(\textbf{x}_{i+1})+f_1(\textbf{x}_{i+2}) \\
  0 \\
  f_2(\textbf{x}_i)-2f_2(\textbf{x}_{i+1})+f_2(\textbf{x}_{i+2}) \\
\end{array}\right) \\
    &  & +\,d_i\textbf{e}_i
    \end{array}
\end{equation}
and
\begin{equation}\label{17}
    \dot{d}_i\,=\,k_i\,\|\textbf{e}_i\|_2^2\,,
\end{equation}
$f_1(\textbf{x}_i)\,=\,a(x_{i2}\,-\,x_{i1}),\,f_2(\textbf{x}_i)\,=\,
x_{i1}x_{i2}\,-\,bx_{i3}$,
$x_{51}\,\equiv\,x_1,\,x_{52}\,\equiv\,x_2$, and
$1\,\leq\,i\,\leq\,50$.

Obviously, one gets
$$
\begin{array}{ccc}
  \bar{\textbf{g}}(\textbf{x}_i,\,s,\,t)
  & = & \left(\begin{array}{c}
  0 \\
  -x_{i1}x_{i3}\,+\,s_1s_3 \\
  x_{i1}x_{i2}\,-\,s_1s_2 \\
\end{array}\right) \\
   & = & \left(\begin{array}{c}
  0 \\
  -x_{i3}e_{i1}\,-\,s_1e_{i3} \\
  x_{i2}e_{i1}\,+\,s_1e_{i2} \\
\end{array}\right)\,, \\
\end{array}
$$
where $1\,\leq\,i\,\leq\,50$.

Since Lorenz attractor is confined to a bounded region
$\Phi\,\subset\,\textbf{R}^3$ [9-13], there exists a constant $M$
satisfying $|x_{ij}|,\,|s_{j}|\,\leq\,M$ for $1\,\leq\,i\,\leq\,50$
and $j\,=\,1,\,2,\,3$. Therefore,
$\|\bar{\textbf{g}}(\textbf{x}_i,\,s,\,t)\|_2\,=\,\sqrt{(x_{i3}e_{i1}\,
+\,s_1e_{i3})^2\,+\,(x_{i2}e_{i1}\,+\,s_1e_{i2})^2}\,\leq\,2M\|\textbf{e}_i\|_2$.

Similarly, one has
$$
\bar{\textbf{h}}_i(\textbf{x}_1,\textbf{x}_2,\cdots,\textbf{x}_N,
\textbf{s})=\left(\begin{array}{c}
  f_1(\textbf{e}_i)-2f_1(\textbf{e}_{i+1})+f_1(\textbf{e}_{i+2}) \\
  0 \\
  f_3(\textbf{x}_i,\,\textbf{x}_{i+1},\,\textbf{x}_{i+2},\,\textbf{s}) \\
\end{array}\right),
$$
where
$f_3(\textbf{x}_i,\,\textbf{x}_{i+1},\,\textbf{x}_{i+2},\,\textbf{s})\,=\,
-be_{i3}\,+\,2be_{i+1,3}\,-\,be_{i+2,3}\,+\,x_{i2}e_{i1}\,+\,s_1e_{i2}\,-\,
2(x_{i+1,2}e_{i+1,1}\,+\,s_1e_{i+1,2})\,+\,x_{i+2,2}e_{i+2,1}\,+\,s_1e_{i+2,2}$
and $1\,\leq\,i\,\leq\,50$.

\begin{figure}
\begin{center}
\includegraphics[width=8cm]{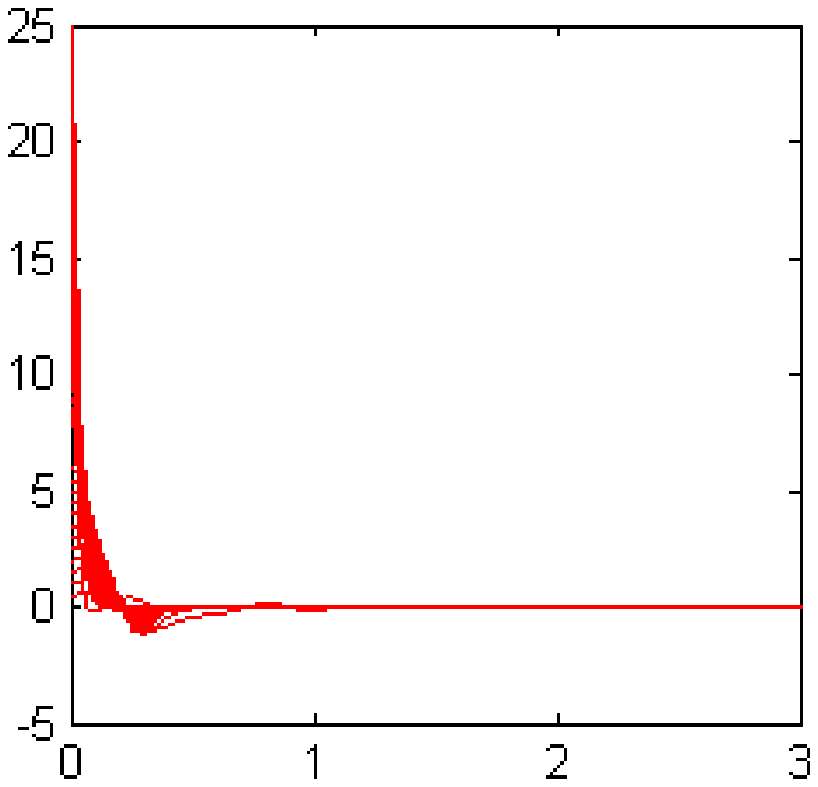}\\
{(a) $\textbf{e}_{i1}$ ($1\,\leq\,i\,\leq\,50$)}\\
\includegraphics[width=8cm]{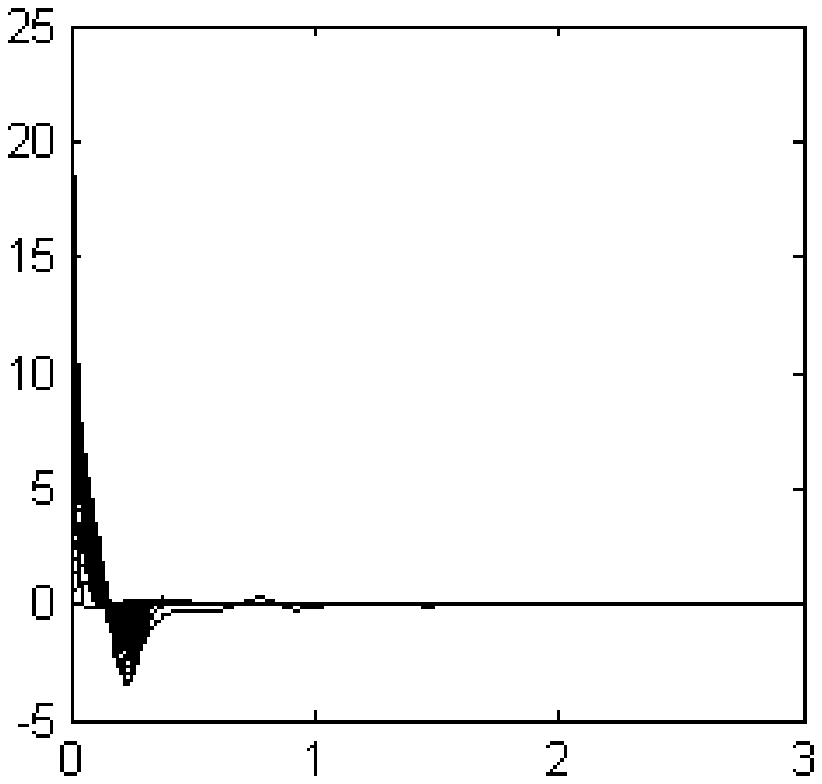}\\
{(b) $\textbf{e}_{i2}$ ($1\,\leq\,i\,\leq\,50$)}\\
\includegraphics[width=8cm]{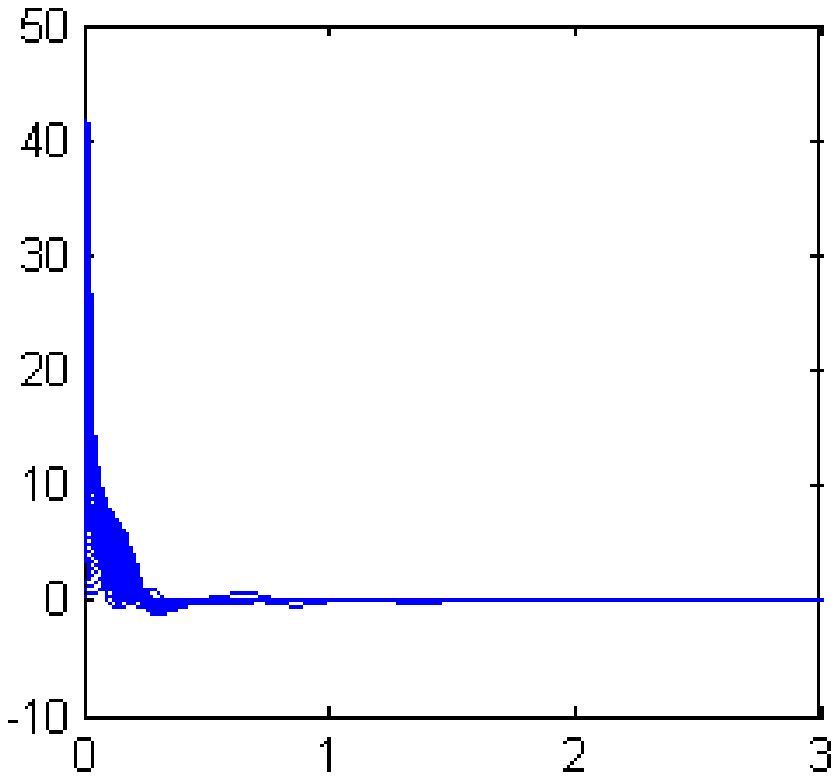}\\
{(c) $\textbf{e}_{i3}$ ($1\,\leq\,i\,\leq\,50$)}\\
{Fig. 1. Synchronization errors of network (\ref{16})-(\ref{17}).}%
\end{center}
\end{figure}

Since
$$
\begin{array}{rl}
   & \|\bar{\textbf{h}}_i(\textbf{x}_1,\textbf{x}_2,\cdots,\textbf{x}_N,\textbf{s})\|_2^2\\
 = &(f_1(\textbf{e}_i)-2f_1(\textbf{e}_{i+1})+f_1(\textbf{e}_{i+2}))^2 \\
   & +\,(f_3(\textbf{x}_i,\,\textbf{x}_{i+1},\,\textbf{x}_{i+2},\,\textbf{s}))^2 \\
 \leq & (a\|\textbf{e}_i\|_1\,+\,2a\|\textbf{e}_{i+1}\|_1\,+\,a\|\textbf{e}_{i+2}\|_1)^2 \\
      & +\,(M\|\textbf{e}_i\|_1\,+\,2M\|\textbf{e}_{i+1}\|_1\,+\,M\|\textbf{e}_{i+2}\|_1)^2 \\
 \leq & 6(a^2\,+\,M^2)(\|\textbf{e}_i\|_1^2\,+\,\|\textbf{e}_{i+1}\|_1^2\,+\,\|\textbf{e}_{i+2}\|_1^2)\\
 \leq & 18(a^2\,+\,M^2)(\|\textbf{e}_i\|_2^2\,+\,\|\textbf{e}_{i+1}\|_2^2\,+\,\|\textbf{e}_{i+2}\|_2^2)\\
 \leq & 18(a^2\,+\,M^2)(\|\textbf{e}_i\|_2\,+\,\|\textbf{e}_{i+1}\|_2\,+\,\|\textbf{e}_{i+2}\|_2)^2\,,\\
\end{array}
$$
where $1\,\leq\,i\,\leq\,50$, one gets
$$
\begin{array}{rl}
  & \|\bar{\textbf{h}}_i(\textbf{x}_1,\textbf{x}_2,\cdots,\textbf{x}_N,\textbf{s})\|_2 \\
  \leq & 3\sqrt{2(a^2\,+\,M^2)}(\|\textbf{e}_i\|_2\,+\,\|\textbf{e}_{i+1}\|_2\,+\,\|\textbf{e}_{i+2}\|_2)\,. \\
\end{array}
$$

Thus H2 and H3 hold. According to Theorem 2, the synchronous
solution $\textbf{S}(t)$ of dynamical network (\ref{16})-(\ref{17})
is globally asymptotically stable.

Assume that
$k_i\,=\,1,\,d_i(0)\,=\,1,\textbf{x}_i(0)\,=\,(4\,+\,0.5i,\,5\,+\,0.5i,\,6\,+\,0.5i)$
for $1\,\leq\,i\,\leq\,50$ and $\textbf{s}(0)\,=\,(4,\,5,\,6)$. The
synchronous error $\textbf{e}_i$ is shown in Fig. 1. Obviously, the
zero error is globally asymptotically stable for dynamical network
(\ref{16})-(\ref{17}).

\smallskip %
{\it Remark 2:} It is well known that the nearest-neighbor coupled
ring lattices are very hard to synchronize. This is because the
coupling coefficient $c$ satisfies $c\,=\,O(N^2)$. However, the
above example shows that the synchronization of nearest-neighbor
coupled ring lattice will be relatively easy by adding a simple
adaptive controller.

\section{Conclusions}

We have further studied the locally and globally adaptive
synchronization of an uncertain complex dynamical network. Several
novel network synchronization criteria have been proved by using
Lyapunov stability theory. Compared with some similar results, our
assumptions and adaptive controllers are very simple. Furthermore,
the effectiveness of these synchronization criteria have been
demonstrated by numerical simulations.

\end{document}